\documentclass[onecolumn,unpublished]{quantumarticle}
\usepackage{amsmath}
\usepackage{amssymb}
\usepackage{amsthm}
\usepackage{amsfonts}
\usepackage[caption=false]{subfig}
\usepackage[colorlinks]{hyperref}
\usepackage[all]{hypcap}
\usepackage{tikz}
\usepackage{color,soul}
\usepackage[utf8]{inputenc}
\usepackage{capt-of}
\usepackage{multirow}
\usepackage{tabularx}
\usepackage{multicol}
\usepackage[numbers]{natbib}

\theoremstyle{definition}

\theoremstyle{definition}

\theoremstyle{definition}

\newcommand{\hide}[1]{}

\DeclareFixedFont{\ttb}{T1}{txtt}{bx}{n}{12} 
\DeclareFixedFont{\ttm}{T1}{txtt}{m}{n}{12}  

\usepackage{color}
\definecolor{deepblue}{rgb}{0,0,0.5}
\definecolor{deepred}{rgb}{0.6,0,0}
\definecolor{deepgreen}{rgb}{0,0.5,0}

\usepackage{listings}

\newcommand\pythonstyle{\lstset{
language=Python,
basicstyle=\ttfamily,
otherkeywords={self,controlled_by,with,Mod,Quint,LookupTable,assert},
keywordstyle=\bfseries\color{deepblue},
commentstyle=\color{gray},
emph={measure,__init__},
emphstyle=\bfseries\color{deepblue},
stringstyle=\color{deepgreen},
showstringspaces=false
}}

\lstnewenvironment{python}[1][]
{
\pythonstyle
\lstset{#1}
}
{}

\renewcommand\thesection{\arabic{section}}

\newcommand{\eq}[1]{\hyperref[eq:#1]{Equation~\ref*{eq:#1}}}
\renewcommand{\sec}[1]{\hyperref[sec:#1]{Section~\ref*{sec:#1}}}
\DeclareRobustCommand{\app}[1]{\hyperref[app:#1]{Appendix~\ref*{app:#1}}}
\newcommand{\fig}[1]{\hyperref[fig:#1]{Figure~\ref*{fig:#1}}}
\newcommand{\tbl}[1]{\hyperref[tbl:#1]{Table~\ref*{tbl:#1}}}
\newcommand{\theoremref}[1]{\hyperref[theorem:#1]{Theorem~\ref*{theorem:#1}}}
\newcommand{\definitionref}[1]{\hyperref[definition:#1]{Theorem~\ref*{definition:#1}}}

\newcommand{\igate}[1]{*+<.6em>{#1} \POS ="i","i"+UR;"i"+UL **\dir{-};"i"+DL **\dir{-};"i"+DR **\dir{-};"i"+UR **\dir{-},"i"}
\newcommand{\imultigate}[2]{*+<1em,.9em>{\hphantom{#2}} \POS [0,0]="i",[0,0].[#1,0]="e",!C *{#2},"e"+UR;"e"+UL **\dir{-};"e"+DL **\dir{-};"e"+DR **\dir{-};"e"+UR **\dir{-},"i"}
\newcommand{\ighost}[1]{*+<1em,.9em>{\hphantom{#1}}}

\newcommand{\pluseq}{\mathrel{+}=}

\newcommand{\timeseq}{\mathrel{\ast}=}

\input{qcircuit}

\begin{document}
\title{Windowed quantum arithmetic}

\date{\today}
\author{Craig Gidney}
\email{craiggidney@google.com}
\affiliation{Google Inc., Santa Barbara, California 93117, USA}

\begin{abstract}
We demonstrate a technique for optimizing quantum circuits that is analogous to classical windowing.
Specifically, we show that small table lookups can allow control qubits to be iterated in groups instead of individually.
We present various windowed quantum arithmetic circuits, including a windowed modular exponentiation with nested windowed modular multiplications, which have lower Toffoli counts than previous work at register sizes ranging from tens of qubits to thousands of qubits.
\end{abstract}

\maketitle

\section{Introduction}
\label{sec:introduction}

In classical computing, a widely used technique for reducing operation counts is ``windowing"; merging operations together by using lookup tables.
For example, fast software implementations of CRC parity check codes process multiple bits at a time using precomputed tables \cite{perez1983crcbyte}.

In this paper, we show that windowing is also useful in quantum computing.
There are situations where several controlled operations can be merged into a single operation acting on a value produced by a small QROM lookup \cite{babbush2018} (hereafter just ``table lookup").

A simple example of a quantum windowing optimization is, when starting a modular exponentiation, look up the final result of the first twenty iterations of the repeated squaring process instead of actually performing those iterations.
At first glance, this may seem like a bad idea.
This optimization saves twenty multiplications, but generating the lookup circuit is going to require classically computing all $2^{20}$ possible results and so will take millions of multiplications.
But physical qubits are noisy \cite{schroeder2009dram,Bare13,Kim2014}, and quantum error correction is expensive \cite{fowler2012surfacecodereview, campbell2018constraintsatisfaction}.
Unless huge advances are made on these problems, fault tolerant quantum computers will be at least a billion times less efficient than classical computers on an operation-by-operation basis.
Given the current state of the art, trading twenty quantum multiplications for a measly million classical multiplications is a great deal.
The cost of performing the quantum lookup is far more significant than the classical multiplications.

In this paper, we present several other examples of using table lookups to reduce the Toffoli count of operations.
The content is organized as follows.
In \sec{background}, we review background information on performing table lookups over classical data addressed by quantum data.
In \sec{results} we present our methods and results.
We provide pseudo code of constructions using table lookups to accelerate several quantum arithmetic tasks related to multiplications involving classical constants.
We compare the cost of windowed multiplication, schoolbook multiplication, and Karatsuba multiplication routines.
We also show how to construct nested lookup optimizations, by performing windowed multiplications inside a windowed exponentiation.
Finally, we summarize our contributions in \sec{conclusion}.

\begin{figure}[h!]
    \centering
    \resizebox{\linewidth}{!}{
        \includegraphics{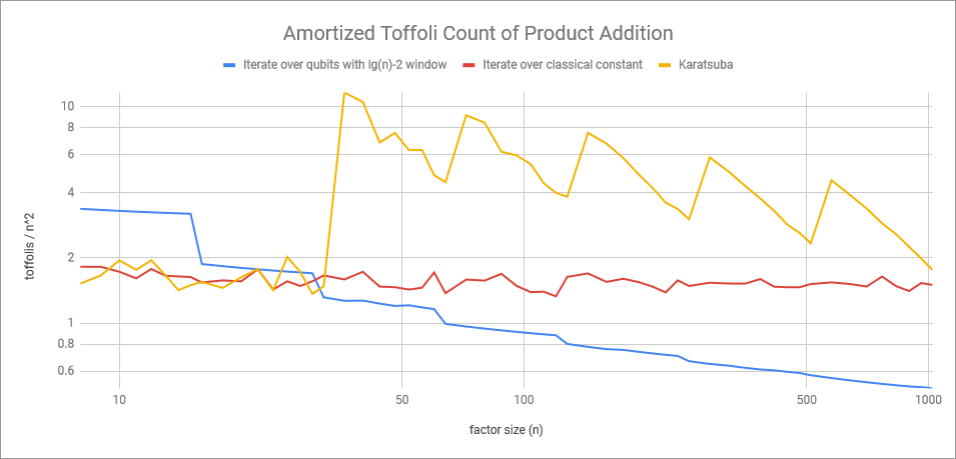}
    }
    \resizebox{\linewidth}{!}{
        \includegraphics{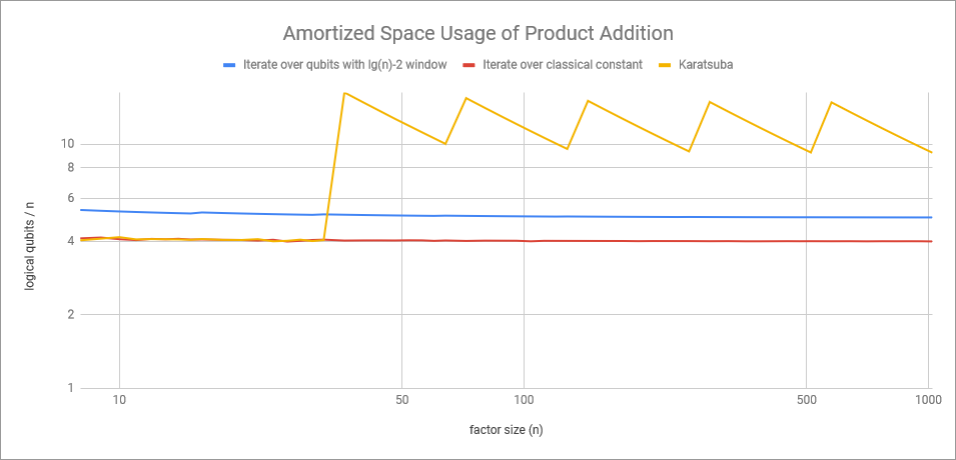}
    }
    \caption{
        \label{fig:product-add-costs}
        Log-log plots of amortized costs of performing the operation $x \pluseq ky$ where $k$ is a classical constant using various constructions.
        Compares windowed multiplication, multiplication via iteration over the classical constant (schoolbook multiplication), and the Karatsuba multiplication procedure from \cite{gidney2019karatsuba} which uses a word size of 32.
        We implemented each of the three methods in Q\# (see ancillary files), and extracted costs using Q\#'s tracing simulator.
        Windowed arithmetic uses slightly more space but has a lower Toffoli count at sizes relevant in practice.
        Karatsuba multiplication will eventually have a lower Toffoli count as the problem size increases, but currently the space overhead is substantial.
    }
\end{figure}

\section{Background: table lookups}
\label{sec:background}

\begin{figure}
\centering
\resizebox{\textwidth}{!}{
\Qcircuit @R=0.7em @C=0.5em {
&\multigate{2}{\text{Input }a}           &\qw& &&&&&&\ctrlo{3} &\qw      &\qw       &\qw      &\qw      &\qw       &\qw      &\qw       &\qw       &\qw      &\qw      &\qw       &\qw      &\qw      &\qw      &\qw       &\qw       &\qw      &\qw      &\qw       &\qw      &\qw       &\qw       &\qw      &\qw      &\qw       &\qw      &\qw      &\ctrl{3} &\qw     &\\
&       \ghost{\text{Input }a}           &\qw& &&&&&&\qw       &\ctrlo{3}&\qw       &\qw      &\qw      &\qw       &\qw      &\qw       &\qw       &\qw      &\qw      &\qw       &\qw      &\ctrl{3} &\qw      &\ctrlo{3} &\qw       &\qw      &\qw      &\qw       &\qw      &\qw       &\qw       &\qw      &\qw      &\qw       &\qw      &\ctrl{3} &\qw      &\qw     &\\
&       \ghost{\text{Input }a}           &\qw& &&&&&&\qw       &\qw      &\ctrlo{3} &\qw      &\qw      &\qw       &\ctrl{3} &\qw       &\ctrlo{3} &\qw      &\qw      &\qw       &\ctrl{3} &\qw      &\qw      &\qw       &\ctrlo{3} &\qw      &\qw      &\qw       &\ctrl{3} &\qw       &\ctrlo{3} &\qw      &\qw      &\qw       &\ctrl{3} &\qw      &\qw      &\qw     &\\
&       \ctrl{4}   \qwx                  &\qw& &&&&&&\ctrl{1}  &\qw      &\qw       &\qw      &\qw      &\qw       &\qw      &\qw       &\qw       &\qw      &\qw      &\qw       &\qw      &\qw      &\ctrl{1} &\qw       &\qw       &\qw      &\qw      &\qw       &\qw      &\qw       &\qw       &\qw      &\qw      &\qw       &\qw      &\qw      &\ctrl{1} &\qw     &\\
&                                        &   & &&&&&&          &\ctrl{1} &\qw       &\qw      &\qw      &\qw       &\qw      &\ctrl{1}  &\qw       &\qw      &\qw      &\qw       &\qw      &\ctrl{1} &\targ    &\ctrl{1}  &\qw       &\qw      &\qw      &\qw       &\qw      &\ctrl{1}  &\qw       &\qw      &\qw      &\qw       &\qw      &\ctrl{1} &\qw      &        &\\
&                                        &   & &&=&&&&          &         &\ctrl{1}  &\qw      &\ctrl{1} &\qw       &\ctrl{1} &\targ     &\ctrl{1}  &\qw      &\ctrl{1} &\qw       &\ctrl{1} &\qw      &         &          &\ctrl{1}  &\qw      &\ctrl{1} &\qw       &\ctrl{1} &\targ     &\ctrl{1}  &\qw      &\ctrl{1} &\qw       &\ctrl{1} &\qw      &         &        &\\
&                                        &   & &&&&&&          &         &          &\ctrl{6} &\targ    &\ctrl{6}  &\qw      &          &          &\ctrl{6} &\targ    &\ctrl{6}  &\qw      &         &         &          &          &\ctrl{6} &\targ    &\ctrl{6}  &\qw      &          &          &\ctrl{6} &\targ    &\ctrl{6}  &\qw      &         &         &        &\\
&\multigate{5}{\oplus\text{Lookup }T_a}     &\qw& &&&&&&\qw       &\qw      &\qw       &\targ^?  &\qw      &\targ^?   &\qw      &\qw       &\qw       &\targ^?  &\qw      &\targ^?   &\qw      &\qw      &\qw      &\qw       &\qw       &\targ^?  &\qw      &\targ^?   &\qw      &\qw       &\qw       &\targ^?  &\qw      &\targ^?   &\qw      &\qw      &\qw      &\qw     &\\
&       \ghost{\oplus\text{Lookup }T_a}     &\qw& &&&&&&\qw       &\qw      &\qw       &\targ^?  &\qw      &\targ^?   &\qw      &\qw       &\qw       &\targ^?  &\qw      &\targ^?   &\qw      &\qw      &\qw      &\qw       &\qw       &\targ^?  &\qw      &\targ^?   &\qw      &\qw       &\qw       &\targ^?  &\qw      &\targ^?   &\qw      &\qw      &\qw      &\qw     &\\
&       \ghost{\oplus\text{Lookup }T_a}     &\qw& &&&&&&\qw       &\qw      &\qw       &\targ^?  &\qw      &\targ^?   &\qw      &\qw       &\qw       &\targ^?  &\qw      &\targ^?   &\qw      &\qw      &\qw      &\qw       &\qw       &\targ^?  &\qw      &\targ^?   &\qw      &\qw       &\qw       &\targ^?  &\qw      &\targ^?   &\qw      &\qw      &\qw      &\qw     &\\
&       \ghost{\oplus\text{Lookup }T_a}     &\qw& &&&&&&\qw       &\qw      &\qw       &\targ^?  &\qw      &\targ^?   &\qw      &\qw       &\qw       &\targ^?  &\qw      &\targ^?   &\qw      &\qw      &\qw      &\qw       &\qw       &\targ^?  &\qw      &\targ^?   &\qw      &\qw       &\qw       &\targ^?  &\qw      &\targ^?   &\qw      &\qw      &\qw      &\qw     &\\
&       \ghost{\oplus\text{Lookup }T_a}     &\qw& &&&&&&\qw       &\qw      &\qw       &\targ^?  &\qw      &\targ^?   &\qw      &\qw       &\qw       &\targ^?  &\qw      &\targ^?   &\qw      &\qw      &\qw      &\qw       &\qw       &\targ^?  &\qw      &\targ^?   &\qw      &\qw       &\qw       &\targ^?  &\qw      &\targ^?   &\qw      &\qw      &\qw      &\qw     &\\
&       \ghost{\oplus\text{Lookup }T_a}     &\qw& &&&&&&\qw       &\qw      &\qw       &\targ^?  &\qw      &\targ^?   &\qw      &\qw       &\qw       &\targ^?  &\qw      &\targ^?   &\qw      &\qw      &\qw      &\qw       &\qw       &\targ^?  &\qw      &\targ^?   &\qw      &\qw       &\qw       &\targ^?  &\qw      &\targ^?   &\qw      &\qw      &\qw      &\qw     &\\
&                                           &   & &&&&&&          &         &          &T_0      &         &T_1       &         &          &          &T_2      &         &T_3 &&&&&&T_4 &&T_5 &&&&T_6 &&T_7   &&&&&&\\
&&&&&&&&& &&& &&&&& &&&&&&& &&& && &&&   &&&&&&\\
}
}
    \caption{
        \label{fig:lookup}
        Table lookup circuit from \cite{babbush2018}.
        The lines emerging from and merging into other lines are AND computations and uncomputations (notation from \cite{gidney2018addition}); they are equivalent to Toffoli gates.
        If the control qubit is set and the address register contains the binary value $a$, then this circuit xors the $a$'th bitstring from a precomputed lookup table $T$ into the $W$ output qubits.
        The diagram is showing the case where $L=2^3$ and $W=6$.
        The question marks beside the CNOT targets indicate that the target should be omitted or included depending on a corresponding bit in $T$.
    }
\end{figure}
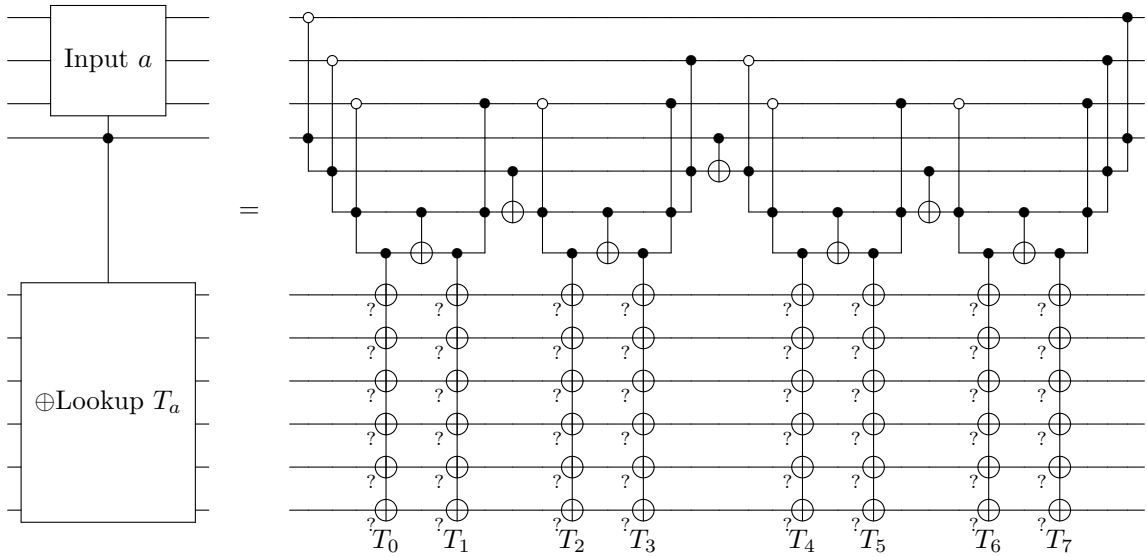

A table lookup is an operation that retrieves data from a classical table addressed by a quantum register.
It performs the operation $\sum_{j=0}^{L-1} |j\rangle|0\rangle \rightarrow \sum_{j=0}^{L-1} |j\rangle|T_j\rangle$, where $T$ is a classically precomputed table with $L$ entries.
In \fig{lookup} we relay a construction that performs this operation with a Toffoli count of $L-1$, independent of the number of bits in each entry, from \cite{babbush2018}.

It is possible to compute a table lookup in $O(WL/k + k)$ Toffolis, where $W$ is the output size of the lookup and $k$ is a freely chosen parameter \cite{low2018qrom}.
Unfortunately, doing so requires $O(Wk)$ ancillae.
The lookups we are performing in this paper are most beneficial when $W$ is large, and so we do not take advantage of this optimization when computing lookups.
However, when uncomputing a lookup, measurement based uncomputation makes these optimizations applicable \cite{berry2019qubitization}.

\begin{figure}
\resizebox{\textwidth}{!}{
\Qcircuit @R=1em @C=0.75em {
  &\qw    &                 \qw&\qw &                               \ctrl{1} &\qw&&& &&&&&\qw    &          \qw       &\qw &     \qw&   \qw&                                     \qw&    \qw&\qw&\qw&               \qw&    \qw&                 \qw&                             \qw&     \qw&                    \ctrl{1}                  &     \qw&                                      \qw&\qw&\\
  &\qw {/}&          \qw       &\qw &       \multigate{1}{\text{Input }a}    &\qw&&& &&&&&\qw {/}&          \qw       &\qw &     \qw&   \qw&                                     \qw&\qw    &\qw&\qw&               \qw&\qw    &          \qw       &                             \qw&     \qw&                   \gate{\text{Input }a_0}    &     \qw&                                      \qw&\qw&\\
  &\qw {/}&\ustick{l}\qw       &\qw &              \ghost{\text{Input }a}    &\qw&&&=&&&&&\qw {/}&\ustick{l}\qw       &\qw &     \qw&   \qw&                                     \qw&\qw    &\qw&\qw&               \qw&\qw    &          \qw       &\gate{\text{Input }a_1}         &     \qw&                                       \qw\qwx&     \qw&             \gate{\text{Input }a_1}     &\qw&\\
  &\qw {/}&\ustick{\geq 2^l}\qw&\qw &        \gate{\text{Unlookup }D_{a}}\qwx& \rstick{\langle0|} \qw   &&& &&&&&\qw {/}&\ustick{\geq 2^l}\qw&\qw &\gate{H}&\meter&                                 \cw    &       &   &   &\lstick{|0\rangle}&\qw {/}&\ustick{2^l}     \qw&\gate{\text{Init Unary }a_1}\qwx&\gate{H}&        \gate{\oplus\text{Lookup }F_{a_0}}\qwx&\gate{H}&        \gate{\text{Clear Unary }a_1}\qwx& \rstick{\langle0|} \qw&    \\
  &       &                    &    &                                        &   &&& &&&&&       &                    &    &        &      &\igate{\text{Compute fixup table }F}\cwx&\cw {/}&\cw&\cw&               \cw&    \cw&                 \cw&                             \cw&     \cw& \cw\cwx&
}
}
    \caption{
        \label{fig:unlookup}
        Uncomputing a table lookup with address space size $L$ and an output size larger than $\sqrt{L}$.
        Has a Toffoli count and measurement depth of $O(\sqrt{L})$.
        Quadratically cheaper than computing the table lookup.
    }
\end{figure}
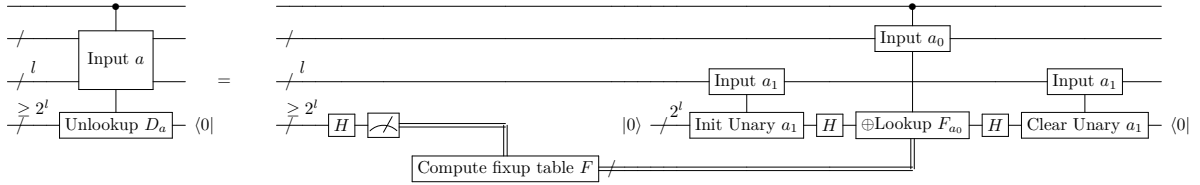

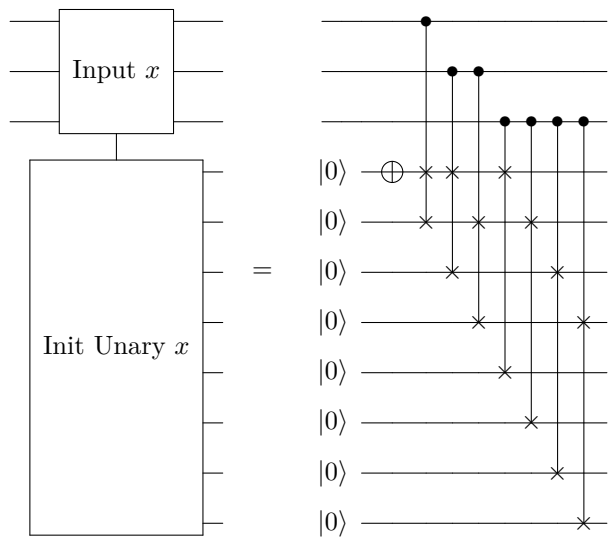
\begin{figure}
\centering
\resizebox{!}{!}{
\Qcircuit @R=1em @C=0.75em {
& \multigate{2}{\text{Input }x}         &\qw && &&& &\qw&\qw                &\qw     &\ctrl{4}&\qw     &\qw     &\qw     &\qw     &\qw     &\qw     &\qw     &\\
&        \ghost{\text{Input }x}         &\qw && &&& &\qw&\qw                &\qw     &\qw     &\ctrl{4}&\ctrl{5}&\qw     &\qw     &\qw     &\qw     &\qw     &\\
&        \ghost{\text{Input }x}         &\qw && &&& &\qw&\qw                &\qw     &\qw     &\qw     &\qw     &\ctrl{5}&\ctrl{6}&\ctrl{7}&\ctrl{8}&\qw     &\\
&\imultigate{7}{\text{Init Unary }x}\qwx&\qw && &&& &   &\lstick{|0\rangle} &\targ   &\qswap  &\qswap  &\qw     &\qswap  &\qw     &\qw     &\qw     &\qw     \\
&       \ighost{\text{Init Unary }x}    &\qw && &&& &   &\lstick{|0\rangle} &\qw     &\qswap  &\qw     &\qswap  &\qw     &\qswap  &\qw     &\qw     &\qw     \\
&       \ighost{\text{Init Unary }x}    &\qw &&=&&& &   &\lstick{|0\rangle} &\qw     &\qw     &\qswap  &\qw     &\qw     &\qw     &\qswap  &\qw     &\qw     \\
&       \ighost{\text{Init Unary }x}    &\qw && &&& &   &\lstick{|0\rangle} &\qw     &\qw     &\qw     &\qswap  &\qw     &\qw     &\qw     &\qswap  &\qw     \\
&       \ighost{\text{Init Unary }x}    &\qw && &&& &   &\lstick{|0\rangle} &\qw     &\qw     &\qw     &\qw     &\qswap  &\qw     &\qw     &\qw     &\qw     \\
&       \ighost{\text{Init Unary }x}    &\qw && &&& &   &\lstick{|0\rangle} &\qw     &\qw     &\qw     &\qw     &\qw     &\qswap  &\qw     &\qw     &\qw     \\
&       \ighost{\text{Init Unary }x}    &\qw && &&& &   &\lstick{|0\rangle} &\qw     &\qw     &\qw     &\qw     &\qw     &\qw     &\qswap  &\qw     &\qw     \\
&       \ighost{\text{Init Unary }x}    &\qw && &&& &   &\lstick{|0\rangle} &\qw     &\qw     &\qw     &\qw     &\qw     &\qw     &\qw     &\qswap  &\qw     \\
}
}
    \caption{
        \label{fig:unary}
        Example circuit producing a unary register from a binary register.
        The qubit at offset $k$ of the unary register will end up set if the binary register is storing $|k\rangle$.
        The general construction, that this circuit is an example of, uses $L$ Fredkin gates (equivalent to $L$ Toffoli gates) where $L$ is the length of the unary register and $n=\lg L$ is the length of the binary register.
        The ``Clear Unary" circuit is this circuit in reverse (and can be optimized using measuring based uncomputation if desired).
    }
\end{figure}

To uncompute a table lookup using measurement based uncomputation, start by measuring all of the output qubits in the X basis.
This will produce random measurement results and negate the phase of random computational basis states in the address register.
However, measuring the qubits frees up significant workspace and the measurement results indicate which computational basis states of the address register were negated.
Fixing the state negations is done using a lookup over a significantly smaller table.
To be specific, the phase negation task is performed by separating the address register into a low half and a high half.
A binary-to-unary conversion (see \fig{unary}) is performed on the low half, and then a table lookup addressed by the high half and targeting the low half is performed.
This creates the opportunity to negate the amplitude of any combination of the computational basis states of the address register.
See \fig{unlookup} for a quantum circuit showing an overview of this process.

The Toffoli count of uncomputing the lookup is $2\sqrt{L}$ instead of $L$.
Uncomputing the lookup has negligible cost, compared to computing the lookup.

\section{Windowed arithmetic constructions}
\label{sec:results}

In this section we will be presenting our windowed arithmetic constructions using pseudo code.
We also provide a few circuit diagrams, but the focus is the code.
Because it is not common to specify quantum algorithms using pseudo code, we will quickly discuss some of the syntactical and semantic choices we've made before continuing.

Although we refer to the snippets as pseudo code, they are actually executable python 3 code.
We have written an experimental python 3 library, available at \href{https://github.com/strilanc/quantumpseudocode}{github.com/strilanc/quantumpseudocode}, which provides the necessary glue.
The snippets are slightly modified versions of code from the  \href{https://github.com/Strilanc/quantumpseudocode/tree/v0.1/examples}{examples folder of the v0.1 release} of that repository.
We do not think this library is ready to be used by others, but we used it to test the code we are presenting and so have made it available.

The basic idea of the pseudo code is that quantum operations are specified in the same way as classical operations, and it is the job of the interpreter to decompose high-level quantum arithmetic into low-level quantum circuit operations.
For example, when $a$ and $b$ are variables holding quantum integers, the statement $a \pluseq b$ applies a quantum addition circuit to $a$ and $b$.
If $b$ is instead a classical integer, then it is treated as a temporary expression (an ``rvalue") that needs to be loaded into a quantum register so that a quantum addition circuit using it can be applied.

There are many other kinds of rvalues that can be temporarily loaded into a register in order to add them into a target.
For example, indexing a lookup table with a quantum integer produces a lookup rvalue and so the statement $a \pluseq T[b]$ results in the following three actions: compute a table lookup with classical data $T$ and quantum address $b$ into a temporary register, then add the temporary register into $a$, then uncompute the table lookup.

We use standard python features such as ranges, slices, and list comprehensions.
We also use a quantum generalization of the ``if c:" block, which we write as ``with controlled\_by(c):" due to technical limitations.
Operations within such a block will be controlled by the qubit ``c".

We use two important custom types: ``Quint" and ``QuintMod".
A Quint is a quantum integer; a register capable of storing a superposition of classical integers.
Every quint has a fixed register length (accessed via ``len"), and stores integers as a sequence of qubits using 2s complement little endian format.
The format is relevant because quints support slicing in order to access subsections of its qubits as a quint.
For example, if ``q" is a 32-qubit quint then ``q[0:8]" is an aliased quint over the least significant byte.
Quints support operations such as addition, subtraction, comparisons, and xoring.
A QuintMod is a quantum integer associated with some modulus.
An inline addition into a modular quint will be performed using modular arithmetic circuits, instead of using 2s complement arithmetic circuits used on quints.

Quantum registers can be allocated and deallocated using ``qalloc" and ``qfree" methods.

\subsection{Product addition}

A product addition operation performs $x \pluseq ky$ where $x$ is a fixed width 2s complement register.
We focus on the case where $k$ is a classical constant.

Normally an implementation of this operation would iterate over the bits of $k$, because this creates many small opportunities for optimization:

\begin{python}
    def plus_equal_product(target: Quint, k: int, y: Quint):
        for i in range(k.bit_length()):
            if (k >> i) & 1:
                target[i:] += y
\end{python}

However, we will instead start from an implementation iterating over the qubits of $y$.
This implementation performs quantumly controlled additions instead of classically controlled additions:

\begin{python}
    def plus_equal_product(target: Quint, k: int, y: Quint):
        for i in range(len(y)):
            with controlled_by(y[i]):
                target[i:] += k
\end{python}

Adding $k$ into $x$ controlled by a qubit $q$ is equivalent to adding into $x$ the result of a table lookup with $q$ as the address, the value $0$ at address 0, and the value $k$ at address 1.
So the above code is equivalent to the following code:

\begin{python}
    def plus_equal_product(target: Quint, k: int, y: Quint):
        table = LookupTable([0, k])
        for i in range(len(y)):
            target[i:] += table[y[i]]
\end{python}

Instead of performing a lookup over one qubit, we can perform a lookup over many qubits.
That is to say, we can introduce windowing:

\begin{python}
    def plus_equal_product(target: Quint,
                           k: int,
                           y: Quint,
                           window: int):
        table = LookupTable([
            i*k
            for i in range(2**window)
        ])
        for i in range(0, len(y), window):
            target[i:] += table[y[i:i+window]]
\end{python}

This windowed implementation of product addition has an asymptotic Toffoli count of $O(\frac{n}{w} (n + 2^w))$ where $w$ is the window size.
Setting the window size to $w=\lg n$, so that the table lookup is as expensive as the addition, achieves a Toffoli complexity of $O(n^2/\lg n)$.

Windowing achieves a logarithmic factor improvement over the construction we started with.
And we show in \fig{product-add-costs} that the advantage is not just asymptotic.
Windowed multiplication has a lower Toffoli count than schoolbook multiplication at relatively small register sizes.

\subsection{Multiplication}

A multiplication operation performs $x \timeseq k$ where $k$ is odd and $x$ is a fixed width 2s complement register.
Because qubits later in $x$ cannot affect qubits earlier in $x$, this operation can be implemented by iterating over $x$, from most significant qubit to least significant qubit, performing controlled additions targeting the rest of the register:

\begin{python}
    def times_equal(target: Quint, k: int):
        assert k % 2 == 1  # Even factors aren't reversible.
        k %= 2**len(target)  # Normalize factor.
        for i in range(len(target))[::-1]:
            with controlled_by(target[i]):
                target[i + 1:] += k >> 1
\end{python}

As in the previous subsection, we can rewrite the controlled addition into a lookup addition, and then window the lookup.
However, there is a complication introduced by the fact that qubits within a window need to operate on each other.
To handle this, we split the inner-loop into two steps: adding the correct value into the rest of the register, and then recursively multiplying within the window.

\begin{python}
    def times_equal(target: Quint, k: int, window: int):
        assert k % 2 == 1  # Even factors aren't reversible.
        k %= 2**len(target)  # Normalize factor.
        if k == 1:
            return
        table = LookupTable([
            (j * k) >> window
            for j in range(2**window)
        ])
        for i in range(0, len(target), window)[::-1]:
            w = target[i:i + window]
            target[i + window:] += table[w]
    
            # Recursively fix up the window.
            times_equal(w, k, window=1)
\end{python}

The Toffoli complexity of this windowed multiplication is $O(\frac{n}{w} \cdot (n + 2^w + w^2))$.
If we set $w = \lg n$ then the Toffoli count is $O(n^2/\lg n)$.

\subsection{Modular product addition}

A modular product addition operation performs $x \pluseq k y \pmod{N}$.
We focus on the case where $k$ and $N$ are classical constants.
We require that $x$, $y$, and $k$ are all non-negative and less than $N$.

Modular product addition is identical to product addition, except that we need to use a modular addition in the inner loop, we need to fold the position factor $2^i$ into the table lookup, and we need to ensure the table lookup returns a value modulo $N$:

\begin{python}
    def plus_equal_product_mod(target: QuintMod,
                               k: int,
                               y: Quint,
                               window: int):
        N = target.modulus
        for i in range(0, len(y), window):
            w = y[i:i + window]
            table = LookupTable([
                j * k * 2**i % N
                for j in range(2**window)
            ])
            target += table[w]
\end{python}

See also \fig{multiply-add}, which generalizes this code to the case where $k$ is a function of a small number of qubits.
This code achieves a Toffoli count of $O(\frac{n}{w} (n + 2^w))$ and, setting $w=\lg n$ as usual, the Toffoli complexity is $O(n^2/\lg n)$.

\subsection{Modular multiplication}

A modular multiplication operation performs $x \timeseq k \pmod{N}$, where $k$ has a multiplicative inverse modulo $N$.
We focus on the case where $k$ and $N$ are classical constants.

Modular multiplication is performed via a series of modular product additions \cite{zalka2006pure, haner2016factoring, gidney2017factoring}, and so this case reduces to the modular product addition case from the previous subsection.
See \fig{multiply}.

\subsection{Modular exponentiation}

A modular exponentiation operation performs $x \timeseq k^e \pmod{N}$, where $k$ has a multiplicative inverse modulo $N$.
We focus on the case where $k$ and $N$ are classical constants.

Modular exponentiation is typically implemented using a series of controlled modular multiplications \cite{vedral1996arithmetic,zalka1998fast,haner2016factoring,gidney2017factoring}.
We can reduce the number of multiplications that are needed by iterating over small windows of the exponent and looking up the corresponding factor to multiply by for each one.
This also removes the need for the multiplications to be controlled, because the table lookup can evaluate to the factor 1 in cases where none of the exponent qubits are set.

There is a catch here.
Windowing over the exponent results in fewer modular multiplications, but the number being multiplied against is now quantum instead of classical.
This could prevent us from applying windowing to the modular multiplications, because windowing isn't faster when both values being multiplied are quantum.
But there is a way around this problem.
Instead of using the exponent qubits to look up the number of multiply by, include the exponent qubits as address qubits in lookups within the windowed modular multiplication.
This allows the intermediate values that are being retrieved to be the correct ones for the factor being multiplied by.

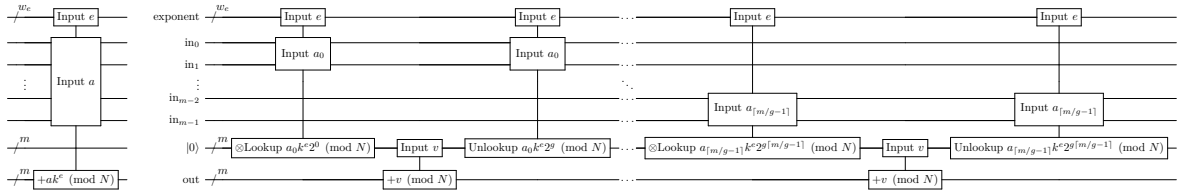
\begin{figure}[h]
\resizebox{\textwidth}{!}{
\Qcircuit @R=1em @C=0.75em {
 \\
 &\qw {/}&\ustick{w_e}\qw&\gate{\text{Input }e}         \qw    &\qw&& &&&&&&&\lstick{\text{exponent}}&\qw {/}&\ustick{w_e}\qw&\gate{\text{Input }e}                           \qw    &                     \qw    &\gate{\text{Input }e}                             \qw    &\qw &\dots &&\gate{\text{Input }e}                                                   \qw    &                     \qw    &\gate{\text{Input }e}                                                                 \qw    &\qw\\
 &\qw    &              \qw&\multigate{4}{\text{Input }a} \qw\qwx&\qw&& &&&&&&&\lstick{\text{in}_0}    &\qw    &          \qw&\multigate{1}{\text{Input }a_0}                 \qw\qwx&                     \qw    &\multigate{1}{\text{Input }a_0}                   \qw\qwx&\qw &\dots &&                                                                        \qw\qwx&                     \qw    &                                                                                  \qw\qwx&\qw\\
 &\qw    &              \qw&       \ghost{\text{Input }a} \qw    &\qw&& &&&&&&&\lstick{\text{in}_1}    &\qw    &          \qw&       \ghost{\text{Input }a_0}                 \qw    &                     \qw    &       \ghost{\text{Input }a_0}                   \qw    &\qw &\dots &&                                                                        \qw\qwx&                     \qw    &                                                                                  \qw\qwx&\qw\\
 &       &           \vdots&                                     &   && &&&&&&&\lstick{\vdots}         &       &             &                                                   \qwx&                            &                                                     \qwx&    &\ddots&&                                                                           \qwx&                            &                                                                                     \qwx&   \\
 &\qw    &              \qw&       \ghost{\text{Input }a} \qw    &\qw&& &&&&&&&\lstick{\text{in}_{m-2}}&\qw    &          \qw&                                                \qw\qwx&                     \qw    &                                                  \qw\qwx&\qw &\dots &&\multigate{1}{\text{Input }a_{\lceil m/g-1\rceil}}                      \qw\qwx&                     \qw    &\multigate{1}{\text{Input }a_{\lceil m/g-1 \rceil}}                               \qw\qwx&\qw\\
 &\qw    &              \qw&       \ghost{\text{Input }a} \qw    &\qw&& &&&&&&&\lstick{\text{in}_{m-1}}&\qw    &          \qw&                                                \qw\qwx&                     \qw    &                                                  \qw\qwx&\qw &\dots &&       \ghost{\text{Input }a_{\lceil m/g-1\rceil}}                      \qw    &                     \qw    &       \ghost{\text{Input }a_{\lceil m/g-1 \rceil}}                               \qw    &\qw\\
 &\qw {/}&\ustick{m}    \qw&                              \qw\qwx&\qw&& &&&&&&&\lstick{|0\rangle}      &\qw {/}&\ustick{m}\qw&\gate{\otimes \text{Lookup }a_0 k^e 2^0\pmod{N}}\qw\qwx&\gate{\text{Input }v}\qw    &\gate{\text{Unlookup } a_0 k^e 2^g \pmod{N}}\qw\qwx&\qw &\dots &&\gate{\otimes \text{Lookup } a_{\lceil m/g-1\rceil} k^e 2^{g \lceil m/g-1\rceil} \pmod{N}}\qw\qwx&\gate{\text{Input }v}\qw    &\gate{\text{Unlookup } a_{\lceil m/g-1\rceil} k^e 2^{g \lceil m/g - 1 \rceil} \pmod{N}}\qw\qwx&\qw\\
 &\qw {/}&\ustick{m}    \qw&\gate{+ak^e \pmod{N}}  \qw\qwx&\qw&& &&&&&&&\lstick{\text{out}}     &\qw {/}&\ustick{m}\qw&                                                \qw    &\gate{+v \pmod{N}}            \qw\qwx&                                                  \qw    &\qw &\dots &&                                                                        \qw    &\gate{+v \pmod{N}}            \qw\qwx&                                                                                  \qw    &\qw\\
 \\
}
}
    \caption{
        \label{fig:multiply-add}
        A windowed modular product addition circuit using windowed arithmetic, where the factor to multiply by is derived from a small number of input qubits from an exponent in a modular exponentiation.
    }
\end{figure}

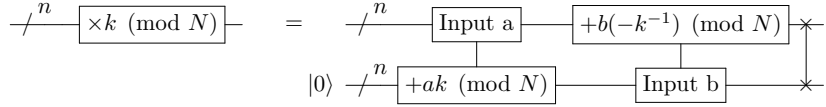
\begin{figure}[h]
\centering
\resizebox{0.7\textwidth}{!}{
\Qcircuit @R=1em @C=0.75em {
 \\
 &\qw {/}&\ustick{n}\qw&\qw &\gate{\times k \pmod{N}}  &\qw&&&=&&&   &\qw {/}&\ustick{n}\qw&\gate{\text{Input a}}&\gate{+b(-k^{-1}) \pmod{N}}          &\qswap    &\qw\\
 &       &             &    &                          &   &&& &&&\lstick{|0\rangle}     &\qw {/}&\ustick{n}\qw&\gate{+ak \pmod{N}}       \qwx&\gate{\text{Input b}}\qwx&\qswap\qwx&\qw\\
 \\
}
}
    \caption{
        \label{fig:multiply}
        Modular multiplication decomposes into modular product additions.
    }
\end{figure}

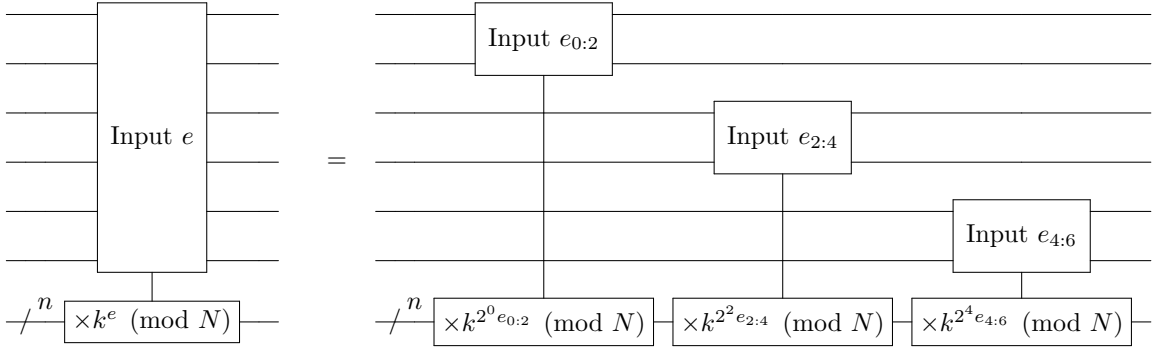
\begin{figure}[h]
\resizebox{\textwidth}{!}{
\Qcircuit @R=1em @C=0.75em {
\\
&\qw&\qw&\multigate{5}{\text{Input }e}&\qw&\qw &&&&& &\qw&\qw&       \multigate{1}{\text{Input }e_{0:2}}&\qw&\qw&\qw&\\
&\qw&\qw&\ghost{\text{Input }e}&\qw&\qw &&&&& &\qw&\qw&              \ghost{\text{Input }e_{0:2}}&\qw&\qw&\qw&\\
&\qw&\qw&\ghost{\text{Input }e}&\qw&\qw &&&&& &\qw&\qw&\qw\qwx&       \multigate{1}{\text{Input }e_{2:4}}&\qw&\qw&\\
&\qw&\qw&\ghost{\text{Input }e}&\qw&\qw &&&=&& &\qw&\qw&\qw\qwx&              \ghost{\text{Input }e_{2:4}}&\qw&\qw&\\
&\qw&\qw&\ghost{\text{Input }e}&\qw&\qw &&&&& &\qw&\qw&\qw\qwx&\qw\qwx&       \multigate{1}{\text{Input }e_{4:6}}&\qw&\\
&\qw&\qw&\ghost{\text{Input }e}&\qw&\qw &&&&& &\qw&\qw&\qw\qwx&\qw\qwx&              \ghost{\text{Input }e_{4:6}}&\qw&\\
&{/}\qw&\ustick{n}\qw&\gate{\times k^e \pmod{N}}\qwx&\qw&\qw &&&&& &{/}\qw&\ustick{n}\qw&\gate{\times k^{2^0 e_{0:2}} \pmod{N}}\qwx& \gate{\times k^{2^2 e_{2:4}} \pmod{N}}\qwx& \gate{\times k^{2^4 e_{4:6}} \pmod{N}}\qwx&\qw&\\
\\
}
}
    \caption{
    \label{fig:exponentiation}
    A six-exponent-qubit modular exponentiation circuit performed using windowed arithmetic with an exponent window size of 2.
    The relevant exponent qubits have to be included as address qubits in lookups within the windowed modular multiplication circuits, as seen in \fig{multiply-add}.
    }
\end{figure}

In the following pseudo code, we have inlined the modular product additions (see \fig{multiply-add}) that are performed as part of the modular multiplications (see \fig{multiply}) that implement the modular exponentiation (see \fig{exponentiation}).
This makes the code longer, but demonstrates how the exponent window and multiplication window are being used together when looking up values to add into registers.

\begin{python}
    def times_equal_exp_mod(target: QuintMod,
                            k: int,
                            e: Quint,
                            e_window: int,
                            m_window: int):
        """Performs `target *= k**e`, modulo the target's modulus."""
        N = target.modulus
        ki = modular_multiplicative_inverse(k, N)
        assert ki is not None
    
        a = target
        b = qalloc_int_mod(modulus=N)
    
        for i in range(0, len(e), e_window):
            ei = e[i:i + e_window]
    
            # Exponent-indexed factors and inverse factors.
            kes = [pow(k, 2**i * x, N)
                   for x in range(2**e_window)]
            kes_inv = [modular_multiplicative_inverse(x, N)
                       for x in kes]
    
            # Perform b += a * k_e (mod modulus).
            # Maps (x, 0) into (x, x*k_e).
            for j in range(0, len(a), m_window):
                mi = a[j:j + m_window]
                table = LookupTable(
                    [(ke * f * 2**j) % N
                     for f in range(2**len(mi))]
                    for ke in kes)
                b += table[ei, mi]
    
            # Perform a -= b * inv(k_e) (mod modulus).
            # Maps (x, x*k_e) into (0, x*k_e).
            for j in range(0, len(a), m_window):
                mi = b[j:j + m_window]
                table = LookupTable(
                    [(ke_inv * f * 2**j) % N
                     for f in range(2**len(mi))]
                    for ke_inv in kes_inv)
                a -= table[ei, mi]
    
            # Relabelling swap. Maps (0, x*k_e) into (x*k_e, 0).
            a, b = b, a
    
        # Xor swap result into correct register if needed.
        if a is not target:
            swap(a, b)
            a, b = b, a
        qfree(b)
\end{python}

We have tested that the above code actually returns the correct result in randomly chosen cases.

The Toffoli complexity of this code is $O(\frac{n_e n}{w_e w_m} (n + 2^{w_e + w_m}))$ where $n_e$ is the number of exponent qubits, $n$ is the register size, $w_e$ is the exponent windowing size, and $w_m$ is the multiplication windowing size.
For the same reason that square fences cover more area per perimeter than rectangular fences, it is best to use roughly even window sizes over the exponentiation and the multiplications.
Setting $w_e=w_m=\frac{1}{2}\lg n$ achieves a Toffoli complexity of $O(\frac{n_e n^2}{\lg^2 n})$, saving two log factors over the naive algorithm.

\subsection{Cost comparison}

We implemented different product addition algorithms in Q\#, and used its tracing simulator to compare their costs.
The results are shown in \fig{product-add-costs}.
The estimation methodology is as follows.
We generated random problems at various sizes.
The problem input at size $n$ is an $n$-bit classical constant, an $n$-qubit offset register, and a $2n$-qubit target register all initialized into a random computational basis state.
At small sizes we sampled several problems, in order to average out noise due to the ``classical iteration" algorithm having a cost that depends on the Hamming weight of the factors.
At larger sizes the Hamming weight varies less (proportionally speaking), so we only sampled single problem instances.
We fed the sampled problems into the three constructions while checking that they each returned the correct result.
We had to omit some optimizations, in particular the table lookup uncomputation optimization, because they were incompatible with Q\#'s Toffoli simulator.

\section{Conclusion}
\label{sec:conclusion}

In this paper we generalized the classical concept of windowing, of using table lookups to reduce operation counts, to the quantum domain.
We presented constructions using this technique for several multiplication tasks involving classical constants.
Although windowed constructions are not asymptotically optimal, e.g. Karatsuba multiplication has a lower asymptotic cost than windowed multiplication, we presented data indicating windowed constructions are more efficient than previous work for register sizes that would be relevant in practice (from tens of qubits to thousands of qubits).

There are cases where windowing does not work.
For example, if all of the registers in $x \pluseq a \cdot b$ are quantum registers, we are not aware of a way to use windowing to optimize the operation.
Windowing should be useful for computing remainders and quotients when the divisor is a classical constant, but we are unsure if it can be used when computing multiplicative inverses modulo a classical constant.
We leave the task of exhaustively surveying which quantum arithmetic tasks benefit from windowing, and which do not, as future work.

Ultimately, windowing is an optimization that improves the cost of several basic quantum arithmetic tasks at practical register sizes.
Because of this, we believe windowing will be a mainstay of quantum software engineering as it has proven to be in classical software engineering.

\section{Acknowledgements}
We thank Ryan Babbush, Adam Langley, Ilya Mironov, and Ananth Raghunathan for reading an early version of this paper and providing useful feedback which improved it.
We thank Austin Fowler, Martin Ekerå, and Johan Håstad for useful feedback and discussions.
We thank Hartmut Neven for creating an environment where this research was possible in the first place.

\bibliographystyle{plainnat}
\bibliography{refs}

\begin{thebibliography}{16}
\providecommand{\natexlab}[1]{#1}
\providecommand{\url}[1]{\texttt{#1}}
\expandafter\ifx\csname urlstyle\endcsname\relax
  \providecommand{\doi}[1]{doi: #1}\else
  \providecommand{\doi}{doi: \begingroup \urlstyle{rm}\Url}\fi

\bibitem[Babbush et~al.(2018)Babbush, Gidney, Berry, Wiebe, McClean, Paler,
  Fowler, and Neven]{babbush2018}
Ryan Babbush, Craig Gidney, Dominic~W Berry, Nathan Wiebe, Jarrod McClean,
  Alexandru Paler, Austin Fowler, and Hartmut Neven.
\newblock Encoding electronic spectra in quantum circuits with linear t
  complexity.
\newblock \emph{Physical Review X}, 8\penalty0 (4):\penalty0 041015, 2018.

\bibitem[Barends et~al.(2014)Barends, Kelly, Megrant, Veitia, Sank, Jeffrey,
  White, Mutus, Fowler, Campbell, Chen, Chen, Chiaro, Dunsworth, Neill,
  O'Malley, Roushan, Vainsencher, Wenner, Korotkov, Cleland, and
  Martinis]{Bare13}
R.~Barends, J.~Kelly, A.~Megrant, A.~Veitia, D.~Sank, E.~Jeffrey, T.~C. White,
  J.~Mutus, A.~G. Fowler, B.~Campbell, Y.~Chen, Z.~Chen, B.~Chiaro,
  A.~Dunsworth, C.~Neill, P.~O'Malley, P.~Roushan, A.~Vainsencher, J.~Wenner,
  A.~N. Korotkov, A.~N. Cleland, and John~M. Martinis.
\newblock Superconducting quantum circuits at the surface code threshold for
  fault tolerance.
\newblock \emph{Nature}, 508:\penalty0 500--503, 2014.
\newblock arXiv:1402.4848.

\bibitem[Berry et~al.(2019)Berry, Gidney, Motta, McClean, and
  Babbush]{berry2019qubitization}
Dominic~W Berry, Craig Gidney, Mario Motta, Jarrod~R McClean, and Ryan Babbush.
\newblock Qubitization of arbitrary basis quantum chemistry by low rank
  factorization.
\newblock \emph{arXiv preprint arXiv:1902.02134}, 2019.

\bibitem[Campbell et~al.(2018)Campbell, Khurana, and
  Montanaro]{campbell2018constraintsatisfaction}
Earl Campbell, Ankur Khurana, and Ashley Montanaro.
\newblock Applying quantum algorithms to constraint satisfaction problems.
\newblock \emph{arXiv preprint arXiv:1810.05582}, 2018.

\bibitem[Fowler et~al.(2012)Fowler, Mariantoni, Martinis, and
  Cleland]{fowler2012surfacecodereview}
A.~G. Fowler, M.~Mariantoni, J.~M. Martinis, and A.~N. Cleland.
\newblock Surface codes: Towards practical large-scale quantum computation.
\newblock \emph{Phys. Rev. A}, 86:\penalty0 032324, 2012.
\newblock URL \url{https://doi.org/10.1103/PhysRevA.86.032324}.
\newblock arXiv:1208.0928.

\bibitem[Gidney(2017)]{gidney2017factoring}
Craig Gidney.
\newblock Factoring with n+ 2 clean qubits and n-1 dirty qubits.
\newblock \emph{arXiv preprint arXiv:1706.07884}, 2017.

\bibitem[Gidney(2018)]{gidney2018addition}
Craig Gidney.
\newblock Halving the cost of quantum addition.
\newblock \emph{Quantum}, 2:\penalty0 74, 2018.

\bibitem[Gidney(2019)]{gidney2019karatsuba}
Craig Gidney.
\newblock Asymptotically efficient quantum karatsuba multiplication.
\newblock \emph{arXiv preprint arXiv:1904.07356}, 2019.

\bibitem[H{\"a}ner et~al.(2016)H{\"a}ner, Roetteler, and
  Svore]{haner2016factoring}
Thomas H{\"a}ner, Martin Roetteler, and Krysta~M Svore.
\newblock Factoring using 2n+ 2 qubits with toffoli based modular
  multiplication.
\newblock \emph{arXiv preprint arXiv:1611.07995}, 2016.

\bibitem[Kim et~al.(2014)Kim, Daly, Kim, Fallin, Lee, Lee, Wilkerson, Lai, and
  Mutlu]{Kim2014}
Y.~Kim, R.~Daly, J.~Kim, C.~Fallin, J.~H. Lee, D.~Lee, C.~Wilkerson, K.~Lai,
  and O.~Mutlu.
\newblock Flipping bits in memory without accessing them: An experimental study
  of dram disturbance errors.
\newblock In \emph{2014 ACM/IEEE 41st International Symposium on Computer
  Architecture (ISCA)}, pages 361--372, June 2014.
\newblock \doi{10.1109/ISCA.2014.6853210}.

\bibitem[Low et~al.(2018)Low, Kliuchnikov, and Schaeffer]{low2018qrom}
Guang~Hao Low, Vadym Kliuchnikov, and Luke Schaeffer.
\newblock Trading t-gates for dirty qubits in state preparation and unitary
  synthesis.
\newblock \emph{arXiv preprint arXiv:1812.00954}, 2018.

\bibitem[Perez(1983)]{perez1983crcbyte}
Aram Perez.
\newblock Byte-wise crc calculations.
\newblock \emph{IEEE Micro}, 3\penalty0 (3):\penalty0 40--50, 1983.
\newblock \doi{10.1109/MM.1983.291120}.

\bibitem[Schroeder et~al.(2009)Schroeder, Pinheiro, and
  Weber]{schroeder2009dram}
Bianca Schroeder, Eduardo Pinheiro, and Wolf-Dietrich Weber.
\newblock Dram errors in the wild: a large-scale field study.
\newblock In \emph{ACM SIGMETRICS Performance Evaluation Review}, volume~37,
  pages 193--204. ACM, 2009.

\bibitem[Vedral et~al.(1996)Vedral, Barenco, and Ekert]{vedral1996arithmetic}
Vlatko Vedral, Adriano Barenco, and Artur Ekert.
\newblock Quantum networks for elementary arithmetic operations.
\newblock \emph{Physical Review A}, 54\penalty0 (1):\penalty0 147, 1996.

\bibitem[Zalka(1998)]{zalka1998fast}
Christof Zalka.
\newblock Fast versions of shor's quantum factoring algorithm.
\newblock \emph{arXiv preprint quant-ph/9806084}, 1998.

\bibitem[Zalka(2006)]{zalka2006pure}
Christof Zalka.
\newblock Shor's algorithm with fewer (pure) qubits.
\newblock \emph{arXiv preprint quant-ph/0601097}, 2006.

\end{thebibliography}

\end{document}